\documentclass[pra,twocolumn]{revtex4}
\usepackage[T1]{fontenc}
\usepackage{graphicx}
\usepackage{hyperref}
\usepackage{ulem}
\usepackage{bbold}
\usepackage[section]{placeins}
\usepackage{amsmath}
\usepackage{amssymb}
\usepackage{dsfont}
\usepackage{color}

\newcommand{\etabar}{\ensuremath{\bar{\eta}}}

\begin{document}

\title{Weak values beyond weak measurement}
\author{Jeff~Z.~Salvail}
\affiliation{Department of Physics, Simon Fraser University, British Columbia, V5A 1S6}
\date{\today}

\begin{abstract}

The AAV effect is the well-known phenomenon where a weak measurement followed by post-selection leads to a pointer shift proportional to the weak value of the measured observable. The effect is usually derived by considering a perturbative expansion of the time-evolution operator corresponding to the measurement.
We show that the AAV effect can be instead derived given three simple conditions on the pointer states. The most important condition is that the pointer states be approximately indistinguishable. The other two conditions relate the observables used to characterize the pointer with the measured system observable.
%%Why is this interesting?
This intuitive approach does not require any model for the measurement process, but is completely consistent with the usual approach based on the von Neumann Hamiltonian.
%%What is new?
Moreover, this approach predicts and explains a new phenomenon, called weak echo, where the usual weak value behaviour can re-emerge at high interaction strengths.
%
%\keywords{Weak Values \and Von Neumann Measurement}
%\PACS{03.65.Ta \and 03.67.-a}
%
\end{abstract}
\maketitle

\section{Introduction}

The weak value was introduced by Aharonov, Albert, and Vaidman (AAV) as part of an effort to understand the behaviour of quantum systems between pre- and post-selection \cite{Aharonov:1988wh}. (See Ref. \cite{Dressel:2014ks} for a pedagogical introduction.) The weak value is, in a sense, a generalization of the standard expectation value. In von Neumann's formulation of measurement, a system and pointer, both quantum mechanical objects, are made to interact in a specific way \cite{VonNeumann:791332}. Characterizing a dynamical variable of the pointer after the interaction reveals information about the system. In the usual case, the pointer's average position is directly proportional to the expectation value of the measured system observable. In this way, it can be said that the system originally possessed the quantity on average, and the measurement revealed this fact.

AAV discovered the weak value by modifying the measurement process in two ways: first, by reducing the strength of the measurement such that the pointer shift is much less than the initial uncertainty of the pointer; second, by post-selecting the system and considering only the conditional pointer pointer shift. AAV found the pointer shift in this case is given by the so-called weak value, rather than the expectation value. This is called the AAV effect. While the expectation value is real and bounded by the eigenvalues of the measured observable, the weak value is both unbounded and potentially complex. A complex pointer shift is defined as a shift in two complimentary observables. These properties have led to some controversy about the interpretation of the weak value as a property of the system \cite{Leggett:1989te,Aharonov:1989vv}. Despite this interpretational difficulty, the weak value has been demonstrated to be measurable and useful.

In fact, these two properties are the key aspects of the surprising and useful behaviour of some recent weak value experiments.
That it is unbounded has led to the use of the weak value for measuring small effects; this is called weak value amplification \cite{Ritchie:1991vf,Hosten:2008ih,Dixon:2009cz}.
Because it is in general complex, the weak value can be used to measure conditional quasi-probabilities \cite{Steinberg:1995tj} that inform about quantum paradoxes \cite{RESCH:2004es,LUNDEEN:2009hq}, and allow the direct measurement of quantum wavefunctions and full characterization of quantum states \cite{Lundeen:2011vz,Lundeen:2012db,Salvail:2013bo,Malik:2014bf,Bamber:2014ih}. 

The standard approach to deriving the AAV effect relies on assuming a Hamiltonian between system and pointer of specific form (the same as used by von Neumann in Ref. \cite{VonNeumann:791332}), and a specific initial pointer state. A first-order approximation of the time evolution operator, followed by post-selection and pointer readout then complete the derivation (see section \ref{sec:Overview} for details). However, there are several conceptual issues associated with this approach, which we group into two categories. The primary aim of this letter is to provide a new framework with which to derive the AAV effect that addresses all of these issues.

The first category is concerned with the notion of weakness. The common criterion for weakness (i.e., the validity of the first-order approximation leading to the AAV effect) states that the measurement strength (defined as shift in the position of the pointer by the measurement) should be much less than the width of the pointer (see, e.g., refs. \cite{Aharonov:1988wh,Hosten:2008ih}). However, some weak value experiments \cite{RESCH:2004es,Pryde:2005gk,LUNDEEN:2009hq,Lundeen:2011vz,Kocsis:2011jg,Goggin:2011iw,Malik:2014bf} use two-state (qubit) Hilbert spaces as the pointer (usually photon polarization) rather than a continuous degree of freedom.
It is not immediately obvious how to connect the initial state uncertainty with measurement strength in the case of a qubit, since the state is usually thought of as a point on (or in) the Bloch sphere and hence the state has no width.
Nor is it obvious whether identical conditions for weakness apply for both real and imaginary components of the AAV effect.
Furthermore, some recent studies report the predictions of weak-value-like behaviour beyond the usual regime of weakness \cite{Dressel:2012ic,Kedem:2010jj}. Clearly a new notion of weakness is needed which can settle all these issues.

The second category is concerned with the notion of measurement. The usual derivation is fundamentally Hamiltonian-centric, as it relies on the specific form of the time-evolution operator. This is not necessarily a problem, but how do we account for weak-value-like behaviour observed in situations with different interactions? While many measurements follow a von Neumann framework, not all do (see the argument of Leggett \cite{Leggett:1989te}). The modular value \cite{Kedem:2010jj} is an example where the measurement Hamiltonian is slightly modified but a weak value can still be observed. A recent experiment \cite{Shomroni:2013fm} demonstrates the AAV effect despite the absence of any von Neumann (or similar) coupling Hamiltonian. An understanding of the separate roles of the von Neumann measurement scheme and weakness would be helpful in explaining these results.

In this Letter, we develop a framework with which to understand the AAV effect that can resolve the issues described above. We derive the AAV effect by calculating the expectation value of pointer observables, conditional on a system post-selection, for a general (pure) entangled state. No assumptions are made about the interaction which created the entangled state. Instead, three conditions, given as approximations to quantities which appear in the expression for the general conditional expectation, are identified which reduce two conditional expectations (such as pointer position and momentum) to the weak value, completing the derivation of the AAV effect.

The first condition is that the pointer states be approximately indistinguishable.
The AAV effect is usually understood as resulting from interference, and it is well-known that distinguishability destroys interference.
The connection of pointer state indistinguishability with weak values has been pointed out before (see Ref. \cite{Feizpour:2011bs}), but the aim of this paper is to place pointer state indistinguishability explicitly as the central condition for the AAV effect. The second and third conditions relate the pointer readout observables with the measured system observable. Importantly, the three conditions are phrased in terms of pointer states and observables, making them general to all types of quantum systems (any dimension or physical manifestation) and easy (in principle) to implement in the lab. Together, the three conditions are referred to as the weak value approximations.

The rest of this paper is organized as follows. In section \ref{sec:Overview} we derive the AAV effect in the usual way using the von Neumann measurement scheme, to establish a consistent notation and discuss the issues we wish to address. In section \ref{sec:Approximations} the AAV effect is derived without appealing to any entangling operation, by showing how the weak value approximations lead to the effect. The remainder of the paper consists of examples that illustrate the role of the weak value approximations. In section \ref{sec:Examples} we use the von Neumann measurement model to calculate exact pointer state overlaps in a typical measurement scenario for three types of pointers, and use them to obtain shift in the regime of approximate indistinguishability and compare with weak values that each measurement is approximating. Unambiguous conditions for the validity of the weak value conditions are determined. In section \ref{sec:Weakechoes}, the two key properties of the weak value, amplification and complexity, are demonstrated for far beyond the regime usually considered as ``weak'', leading to a phenomenon which we refer to as weak echo. The weak echo relies on the use of periodic or quasi-periodic pointers. Interestingly, the experimentally post-selected state may not correspond with the expression for the weak value that is being measured in this case.
%In section \ref{sec:Complimentary}, another example is worked through which shows how in a von Neumann measurement, the AAV effect is not limited to the pointer, and post-selecting the pointer can lead to a manifestation of the AAV effect in the system which we call a complimentary weak value.
Finally in section \ref{sec:Conclusions} we conclude.

\section{Overview of AAV's Scheme} \label{sec:Overview}

In this section we shall briefly discuss the formalism of von Neumann measurement, and the scheme discovered by AAV to directly measure a weak value. Begin with an uncorrelated system-pointer state $|\Psi_i\rangle = |\psi\rangle \otimes |\xi_0\rangle$ in the joint Hilbert space $\mathcal{H} = \mathcal{H}_{s} \otimes \mathcal{H}_{p}$. The system states live in $\mathcal{H}_s$ and the pointer states are represented by vectors in $\mathcal{H}_p$. The system state $|\psi\rangle$ is called the initial system state, about which we wish to learn something by measuring it. The pointer state is $|\xi_0\rangle$, a suitable configuration that the experimenter identifies as the pointer's zero. Assume that $\mathcal{H}_s$ is a discrete Hilbert space, for example representing the spin of a particle, and $\mathcal{H}_p$ continuous, such as the position. AAV used a Gaussian centred at $x=0$ as the initial pointer state. This is the most common, though others may be used such as Lorentzian or, for two-dimensional Hilbert space, a desired initial qubit state \cite{Shomroni:2013fm}.

Let $|\Psi_i\rangle$ evolve unitarily under the measurement Hamiltonian $H = g(t) \hat{A} \otimes \hat{P}$. The action of this Hamiltonian is to measure the system observable $\hat{A}$ by shifting the pointer using $\hat{P} = -i\hbar \partial / \partial x$ as a generator. For simplicity, let $g(t)$ have compact support $\mathcal{M}$ near the origin and define $\eta = \int_{\mathcal{M}} g(t) \mathrm{d} t$. Most authors refer to $\eta$ as the measurement strength, but we will save that term for another quantity introduced in the next section, and call $\eta$ the measurement parameter. The final system-pointer state $|\Psi_f\rangle$ after the measurement $\mathcal{M}$ is 
\begin{align}
|\Psi_f\rangle &= \exp \left( \frac{-i}{\hbar} \int_{\mathcal{M}} H \mathrm{d} t \right) |\Psi_i\rangle \nonumber \\
&= \sum_{\ell=0}^{\infty} \frac{1}{\ell!}\left( \frac{-i \eta}{\hbar} \hat{A} \otimes \hat{P} \right)^{\ell} |\psi\rangle \otimes |\xi_0\rangle. \label{exactsumfinalstate} \\
&= \sum_n c_n |a_n\rangle \otimes |\xi_n\rangle, \label{finalentangledstate}
\end{align}
 where the $|a_n\rangle$ form the orthonormal eigenbasis of $\mathcal{H}_s$ associated with $\hat{A}$, i.e., $\hat{A}|a_i\rangle = a_i |a_i\rangle$. The system state was expanded in this basis to give the complex coefficients $c_n$, i.e., $|\psi\rangle = \sum_n c_n |a_n\rangle$. The pointer states $|\xi_n\rangle$ are shifted versions of the initial pointer state, i.e., $|\xi_n\rangle = U_n |\xi_0\rangle$ with $U_n = \exp (-i \eta a_n  \hat{P}/\hbar)$. Equation (\ref{finalentangledstate}) is exact and this procedure is the von Neumann measurement model.

Proceed by projecting the system onto state $|f\rangle$ corresponding to post-selection on outcome $f$ (e.g., by performing a second, strong measurement using a different pointer). The resulting (unnormalized) pointer state is $|\phi\rangle = \langle f|\Psi_f\rangle$.

AAV then argued that if the measurement parameter $\eta$ is small (the measurement is weak), we can %truncate the sum of equation  (\ref{exactsumfinalstate}) after the first two terms, giving $|\Psi_f\rangle \approx (\mathds{1} - i \eta/\hbar \hat{A} \otimes \hat{P}) |\Psi_i\rangle$.
%
%AAV actually replaced 
%replace the sum equation (\ref{exactsumfinalstate}) with 
write the state $|\phi\rangle$ of the pointer as
\begin{align}
|\phi\rangle &\approx \sum_{\ell=0}^{\infty} \frac{1}{\ell!} \left(-\frac{i \eta}{\hbar} \frac{\langle f | \hat{A} | \psi \rangle}{\langle f | \psi \rangle} \hat{P} \right)^\ell |\xi_0\rangle \label{weakvaluesum}\\
&= \exp \left(-\frac{i\eta}{\hbar} \frac{\langle f | \hat{A} | \psi \rangle}{\langle f | \psi \rangle} \hat{P} \right)|\xi_0\rangle, \label{AAVshift}
\end{align}
% to argue that the weak measurement shifts the initial wavefunction by the weak value. We use the shift in conditional pointer expectations because it more closely aligns with an experimental implementation of a weak value experiment, where the average pointer shift is recorded. The exact expression, AAV's approximation, and the first order truncation are all equivalent to first order in $\eta$. The argument summarized here is due to Lundeen and Resch in Ref. \cite{Lundeen:2005jr}.
This is a slightly shifted version of the initial pointer state wavefunction. From equation \ref{AAVshift}, we can calculate conditional expectation values $\langle \hat{O} \rangle_{p | s = f}$, corresponding to any pointer observable $\hat{O}$, according to the standard expectation value: $\langle \hat{O} \rangle_{p | s = f} = \langle \phi | \hat{O} | \phi \rangle / \langle \phi | \phi \rangle$. It is assumed that $\langle \phi|\phi\rangle \ne 0$, i.e., that post-selection probability is nonzero (the post-selected pointer shift is undefined if no systems are successfully post-selected).

Using the approximate form of $|\phi\rangle$, and keeping terms only up to first order in $\eta$, it can be shown that (for the Gaussian initial pointer state, and appropriate choice of units for $\hat{X}$ and $\hat{P}$, see \cite{Lundeen:2005jr}):
\begin{equation} \label{AAVeffect}
\frac{1}{\eta} \langle \hat{X} + i\hat{P} \rangle_{p | s = f} \approx  \frac{\langle f | \hat{A} | \psi \rangle}{\langle f | \psi \rangle} =: \langle \hat{A} \rangle^W_f.
\end{equation}
Hence the result of the measurement, as indicated by the average pointer shifts, is the weak value. Equation (\ref{AAVeffect}) is the quantitative statement of the AAV effect. Since $\hat{X}$ and $\hat{P}$ do not commute, the experiment must be repeated for two different subensembles to read the real and imaginary parts of the weak value.

This derivation can be modified to manifest the AAV effect with a qubit pointer, by replacing $\hat{P} \to -\sigma_y$ and $\hat{X} \to \sigma_x$. In this case the initial pointer state should be taken as $\left|\downarrow_z \right\rangle$, the second eigenstate of $\sigma_z$ ($\sigma_{x,y,z}$ are the Pauli matrices) \cite{Lundeen:2005jr}.

As mentioned in the previous section, there are several open questions raised by this approach, which we group into two categories.

The first category is concerned with the notion of weakness. What is/are the exact condition(s) for weakness, namely, under what conditions is it appropriate to replace the sum of equation (\ref{exactsumfinalstate}) with that of (\ref{weakvaluesum})? Since (\ref{exactsumfinalstate}) is a matrix Taylor expansion, it is not easy to give a clear condition. Usually, $\eta$ much less than initial pointer state uncertainty is given. It is unclear how universal this is: are the conditions quantitatively the same for both readout observables, i.e., for $\langle \hat{X} \rangle_{p | s = f} \approx \eta \mathrm{Re} \left[ \langle \hat{A} \rangle^W_f \right]$ and $\langle \hat{P} \rangle_{p | s = f} \approx \eta \mathrm{Im} \left[ \langle \hat{A} \rangle^W_f \right]$? It turns out, not in general. What about for qubit pointers, for which it is not so obvious how to relate the initial pointer state uncertainty with pointer shift (which is an angle)? And, how does this picture account for the observation of weak-value-like behaviours outside this regime as reported in Refs. \cite{Kedem:2010jj,Dressel:2012ic}?

The second category is concerned with the notion of measurement. Is the von Neumann Hamiltonian necessary for the observation of the AAV effect? It is already known that the answer is no: The modular value \cite{Kedem:2010jj} is an example where the measurement Hamiltonian is slightly modified but a weak value can still be observed. A recent experiment \cite{Shomroni:2013fm} demonstrates the AAV effect despite the absence of any von Neumann (or similar) coupling Hamiltonian, hence can not really be considered a weak \textit{measurement} at all, in that framework.

In the next section, the AAV effect will be re-derived in a way that can address all of the issues above.

\section{The Weak Value Approximations} \label{sec:Approximations}

In this section, we derive the conditions, given as approximations to quantities that appear in the exact expression, in order for a general entangled system-pointer state and readout observables to yield a weak value. This is a derivation of the AAV effect by considering only properties of the entangled system-pointer state, not based on the interaction which created this state. The intent is not to supplant the usual derivation, but to offer a parallel, equivalent view, so that understanding can be deeper. 

First, a word on the names \textit{system} and \textit{pointer}. In the von Neumann scheme of the previous section these are identifiable by their role in the interaction: the system begins in an arbitrary superposition, the pointer begins in the ``zero'' state; the diagonalizing observable $\hat{A}$ operates on system states, while the generator $\hat{P}$ is a pointer observable. In the present derivation we are careful not to assume anything about the interaction or state to begin with. However, we can still identify the system and pointer by deciding ahead of time which will be subjected to post-selection, and which will be read out: ``system'' refers to the former and ``pointer'' to the latter. %In Section \ref{sec:Complementary} we will see that this choice is made without loss of generality.

We begin by considering a general state $|\Psi\rangle$ in the joint Hilbert space $\mathcal{H}_s \otimes \mathcal{H}_p$ with orthonormal bases $\{|a_n\rangle \}$ of $\mathcal{H}_s$ and $\{|b_m\rangle\}$ of $\mathcal{H}_p$:
\begin{align} \label{postmeasurestate}
|\Psi\rangle = \sum_{n,m} \gamma_{nm}  |a_n\rangle \otimes |b_m\rangle.
\end{align}
This state can be expressed in the form of equation (\ref{finalentangledstate}), by defining $|\xi_n\rangle = 1/c_n \sum_m \gamma_{nm}|b_m\rangle$ with $|c_n|^2 = \sum_m |\gamma_{nm}|^2$. The states $|\xi_n\rangle$ may or may not be orthogonal, and we call them the pointer states, since there is one correlated with each system basis state $|a_n\rangle$. %We could choose a $|\xi_0\rangle$ and define a $U_n$ which takes $|\xi_0\rangle$ to each $|\xi_n\rangle$, though they will not necessarily have the form of the $U_n$'s in the previous section. We will not need to use such unitary operators at all in this section, nor consider any pre-measurement state.

As in the previous section, we can post-select the system and read the pointer shift conditional on this post-selection. Projecting the system onto state $|f\rangle \in \mathcal{H}_s$ corresponds to post-selecting outcome $f$.
The conditioned pointer expectation can be calculated by renormalizing the projected state, followed by calculating the expectation value in the usual way. Writing the (unnormalized) projected state as $|\phi\rangle = \langle f | \Psi\rangle \in \mathcal{H}_p$, we write for arbitrary pointer observable $\hat{O}$ the conditioned expectation $\langle \hat{O}\rangle_{p | s = f} = \langle \phi| \hat{O} | \phi\rangle / \langle \phi|\phi\rangle$. We calculate this exactly:
\begin{align} \label{generalcase}
\langle \hat{O}\rangle_{p | s = f}
&= \frac{\sum_{n,m}  \langle f | \Pi_{a_{n}} \rho  \Pi_{a_{m}} | f\rangle  O_{mn} }{\sum_{n,m} \langle f | \Pi_{a_{n}} \rho \Pi_{a_{m}} | f \rangle  \mathcal{D}_{mn} }, 
\end{align}
where we have written $\rho = |\psi \rangle \langle \psi|$ with $|\psi\rangle = \sum_n c_n|a_n\rangle$, $\Pi_{a_n} = |a_n\rangle \langle a_n|$, $O_{mn} = \langle \xi_m| \hat{O}  |\xi_n \rangle$, and $\mathcal{D}_{mn} = \langle \xi_m |\xi_n \rangle$.  Note that $O_{mn}$ are not the matrix elements of $\hat{O}$ since the pointer states $|\xi_n\rangle$ are not (necessarily) orthogonal.

One obtains a weak value and the AAV effect from Eq. (\ref{generalcase}) in the following way. Substitutions for the pointer quantities $\mathcal{D}_{mn}$ and $O_{mn}$ are made which reduce equation (\ref{generalcase}) to the AAV effect. The substitutions are given in the form of approximations, and the conditions for the validity of these approximations are the weak value conditions.

The first approximation Eq. (\ref{conditions}a) below states that the pointer states be (approximately) indistinguishable. This is the fundamental requirement to measure a weak value. It is now well-established that the AAV effect arises due to interference, and a core result of quantum mechanics is that distinguishable outcomes will not interfere. It is not surprising therefore that the pointer state distinguishability plays an important role in the AAV effect. %We define a measurement's weakness as its propensity to exhibit the AAV effect, which we shall see depends critically on the pointer state overlaps, and hence we call $|\mathcal{D}_{mn}|$ the weakness. A weakness of $|\mathcal{D}_{mn}| \approx 1$  $\forall m,n$ corresponds to a weak measurement, while $\mathcal{D}_{mn} = \delta_{m,n}$ corresponds to orthogonal pointer states and a strong measurement. We will hence introduce $S_{mn} = 1 - |\mathcal{D}_{mn}|$ captures the measurement strength, and $S_{mn} = 1$, $m \ne n$ is a strong measurement. Note that $\mathcal{D}_{mn}$ also carries phase information as it may be complex, and we will investigate this later in section \ref{sec:Weakechoes}.

%Once again we emphasize that the strength and weakness are defined independently of any measurement interaction, and is distinct from the von Neumann measurement parameter $\eta$ of the previous section, and depends only on the entangled state $|\Psi\rangle$.

The second two approximations Eq. (\ref{conditions}b,c) relate the pointer readout observables with the Hermitian operator that will be called the measured observable. Specifically, the spectrum of the observable is determined this way, and the eigenbasis is determined by the choice of system basis $|\Psi\rangle$ was expanded in, i.e., $\{|a_n\rangle\}$. Call the pointer readout variables $\hat{\chi}$ and $\hat{\mu}$, and introduce the real numbers $\alpha_n$, each corresponding to a pointer 	state $|\xi_n\rangle$. Finally, we introduce a constant $\eta \ne 0$ which accounts for units between system and pointer quantities.

The weak value approximations are:
\begin{subequations} \label{conditions}
\begin{align}
\mathcal{D}_{mn} = \langle \xi_m|\xi_n\rangle &\approx 1, \\
\chi_{mn} = \langle \xi_m| \hat{\chi}  |\xi_n \rangle &\approx \eta \frac{\alpha_m + \alpha_n}{2},  \\
\mu_{mn} = \langle \xi_m| \hat{\mu}  |\xi_n \rangle &\approx i \eta \frac{\alpha_m - \alpha_n}{2}.  
\end{align}
\end{subequations}
Calculate $\left\langle \hat{\chi} \right\rangle_{p | s = f}$ using Eqs. (\ref{generalcase}, \ref{conditions}a,b) and calculate $\left\langle \hat{\mu} \right\rangle_{p | s = f}$ using Eqs. (\ref{generalcase}, \ref{conditions}a,c). Write $\hat{A} = \sum_i \alpha_i \Pi_{a_{i}}$ and use the closure of the projectors, to yield the AAV effect
\begin{equation} \label{weakvalue}
\frac{1}{\eta} \left\langle \hat{\chi} + i  \hat{\mu} \right\rangle_{p | s = f} \approx \frac{ \langle f| \hat{A} \rho  |f\rangle }{\langle f| \rho | f \rangle} =  \langle \hat{A} \rangle^W_f.
\end{equation}
Hence, $\hat{\chi}$ reads the real part of the weak value and $\hat{\mu}$ reads out the imaginary part. Equation (\ref{weakvalue}) as a consequence of Eqs. (\ref{generalcase},\ref{conditions}) is the main result of this Letter. 

The derivation of the AAV effect in this way applies to any case where the von Neumann scheme is used, as a small von Neumann measurement parameter will always lead to pointer states that are approximately indistinguishable. This is discussed in more detail in the next section. However, some experiments do not follow this paradigm and must be mapped \textit{a posteriori} to a von Neumann-like model. The recent experiment  by Shomroni \textit{et al.} is a perfect example; to derive their weak value, a fictional time operator was used to generate the pointer states \cite{Shomroni:2013fm}. Of course, as the authors point out, this had no correspondence to the experiment itself, which provides an entangled system-pointer state via the coupling of the atomic ensemble to the environment, as opposed to through an impulsive coupling between two initially independent quantum degrees of freedom. The approach in this section can be used to derive the AAV effect in this and other situations that do not use von Neumann measurement, by considering only the entangled states, and not supposing interaction of specific forms that may or may not have correspondence to reality.

%Relation of pointer state distinguishability to measurement back-action.

\section{Examples} \label{sec:Examples}

In this section, we look at several examples of von Neumann measurements to explore the role of the the weak value approximations above. For three different types of pointer we calculate the $\mathcal{D}_{mn}$ and corresponding pointer shifts. It will be shown that for all three pointers, the AAV effect is predicted to occur in the usual regime (small interaction parameter) and, for the periodic or quasi-periodic pointers, outside this regime leading to the phenomenon we call a weak echo. %Note that Dressel and Jordan \cite{Dressel:2012ic} examined the Hermite-Gaussian modes, which share qualitative features with the behaviour of the optical pulse that we will explore

The role of the von Neumann interaction in this section is to provide a causal link between an initial uncorrelated system-pointer state and the final entangled state. Hence the weak value measured by the pointer after the interaction has a physical correspondence with the initial, pre-measurement system state; in other words, the measurement revealed something about $|\psi\rangle$. The von Neumann interaction also provides, through the measurement parameter $\eta$, a way to tune the final entangled state and examine the effect of pointer state distinguishability on the conditioned pointer shift. We do not Taylor expand the time-evolution unitary operator in this section.

We will take three physical implementations of the measurement pointer: a Gaussian spatial distribution, a few-cycle linearly polarized optical pulse, and a qubit. The Gaussian temporal envelope of a pulse has been examined already for light \cite{Brunner:2004cf,Brunner:2010dt,Strubi:2013hb} and neutrinos \cite{Berry:2011eh}, which is equivalent in our analysis to a Gaussian spatial distribution. We will take the pointer states in each example to be respectively
\begin{subequations}\label{pointers}
\begin{align} 
G_i(x) &= (2 \pi \sigma^2)^{-1/4}  \exp\left(\frac{-(x-\eta \alpha_i )^2} {4\sigma^2}\right), \\
E_i(\tau) &= E_0 \exp\left(\frac{-(\tau-\eta \alpha_i )^2} {4\sigma^2}\right) \cos(\omega (\tau - \eta \alpha_i)), \\
|Q_i\rangle &= \cos (\eta \alpha_i)|1\rangle + \sin (\eta \alpha_i) |0\rangle.
\end{align}
\end{subequations}
In the above, $\sigma$ is the Gaussian's standard deviation (for the optical pulse, of the envelope). For the optical pulse, $E_0$ represents the electric field at time zero. The choice of carrier-envelope phase in this example is arbitrary but should be taken as fixed to maintain coherence at this time scale in the ensemble of photons. In both of these continuous pointers, the parameter $\eta$ represents the shift from $0$ (in position or time) of the pointer state. Finally the qubit state $|Q_1\rangle$ is a rotation of the state $|1\rangle$ by the angle $\eta$.

For simplicity, we will take the measured system for all three examples to be a qubit, and the measured observable to be $\hat{A} = \Pi_1 = |1\rangle \langle 1|$. Hence the $|a_i\rangle$ in our post-measurement state $|\Psi\rangle$ of equation (\ref{finalentangledstate}) are the eigenstates of the operator: $|0\rangle$, $|1\rangle$ with corresponding eigenvalues $\alpha_0 = 0$, $\alpha_1 = 1$ respectively. 

In each case, a unitary operator can be written which takes an initial state $|\psi\rangle\otimes |\xi_0\rangle$ to its respective final state. Section \ref{sec:Overview} shows the derivation of the final state based on a Hamiltonian of von Neumann type. We take $\hat{H} = g(t)\hat{A} \otimes \hat{\mu}$, and $\hat{\mu}$ is the generator for the measurement. To shift the Gaussian we take $\hat{\mu} \to \hat{P} = -i\hbar \partial/ \partial x$; to shift the optical pulse in time, we take $\hat{\mu} \to -E = -i\hbar \partial/ \partial \tau$; to shift the qubit we take $\hat{\mu} \to -\hbar \sigma_y /2$. 
Each time-evolution operator is then, with appropriate substitutions for $\hat{\mu}$:
\begin{equation} \label{vonNeumannunitary}
U = \Pi_0 \otimes \mathds{1} + \Pi_1 \otimes \exp( -i \eta \hat{\mu} / \hbar),
\end{equation}
where, as before, $\eta = \int_{\mathcal{M}} g(t)\mathrm{d}t$. Note that $t$ and $\tau$ are treated independently: $t$ is a the physical time, $\tau$ pointer state abscissa.
 
\begin{figure}[htbp]
\begin{center}
\includegraphics[]{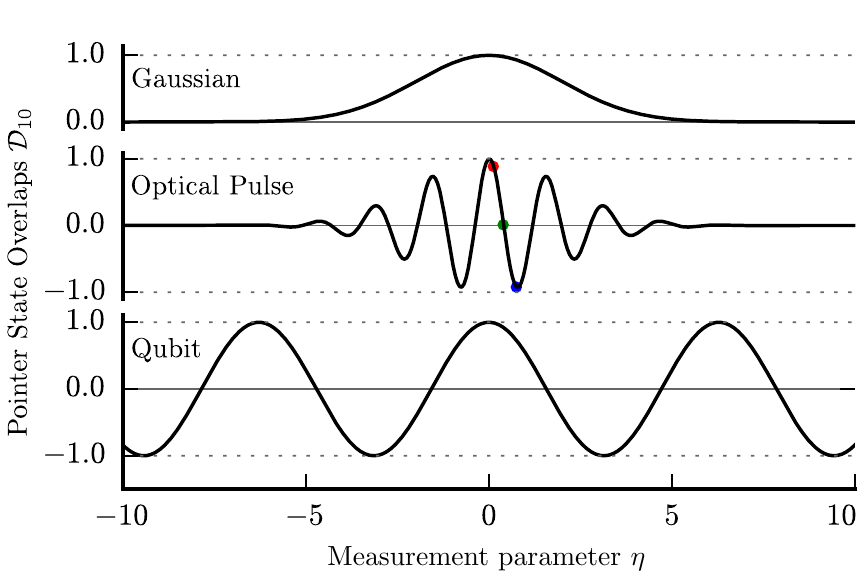}
\caption{\label{D_jis} Pointer state distinguishabilities for Gaussian, optical pulse, and qubit pointers. When $\eta \to 0$, $\mathcal{D}_{10} \to 1$ for all cases, and this is the standard weak regime. When $|\mathcal{D}_{10}| = 0 $, this is a strong measurement, as measurement outcomes are fully resolved. When $\mathcal{D}_{10} \sim 1$ and $\eta \ne 0$, a weak echo can be observed, where weak value behaviour re-emerges. The three coloured disks  indicate values for $\eta$ used in figure \ref{pointershifts_re_im} to calculate state-dependent pointer shifts for the three regimes. Their numerical values are respectively $\eta = (0.12, 0.39, 0.75)$. To calculate these curves, $\sigma = 1$ for the Gaussian and optical pulse, and $\omega = 4$ for the optical pulse. All these curves correspond to the measurement captured in the von Neumann unitary equation (\ref{vonNeumannunitary}), with $\hat{P}$ appropriate to each respective pointer as explained in the text.
}
\end{center}
\end{figure}

First we calculate $\mathcal{D}_{10}$ for the three cases; the results are shown in Fig. \ref{D_jis}. For all three pointers,  the same qualitative behaviour is seen near $\eta = 0$: the curve $\mathcal{D}_{10}(\eta)$ is equal to 1, even in $\eta$, and concave down at this point. The Taylor expansion in $\eta$ for all have the first two terms $1 - c \eta^2$, with $c>0$ being determined by the specifics of the pointer. Hence, for small $\eta$ ($c \eta^2 \ll 1$), we can satisfy the first weak value approximation (\ref{conditions}a).

For the continuous pointers, the two pointer conditions (\ref{conditions}b,c) are satisfied exactly. Use for $\hat{\mu}$ the variables mentioned above for all cases. Use $\hat{\chi} \to \hat{X}$ for the Gaussian and $\hat{\chi} \to \tau$ for the optical pulse. Note that for the two continuous pointers, the $\hat{\chi}$ variable is chosen conjugate to the shift-generating operator in the Hamiltonian; this is in accordance with the Stone-von Neumann theorem (see \cite{Rosenberg:2004tx, Dressel:2014ks}

%Using the Pauli matrix $\hat{\chi} \to \sigma_x$ satisfies (\ref{conditions}b,c) to first order in $\eta$ for the qubit \cite{Lundeen:2005jr}. The qubit pointer is not continuous, so the Stone-von Neumann theorem does not apply, and yet the conditions (\ref{conditions}b,c) are satisfied in qualitatively a similar way, choosing for $\hat{\chi}$ an observable conjugate to the one used in the Hamiltonian. In this case, the weakness plays a crucial role in satisfying these conditions, where it does not for the continuous case.

The Stone-von Neumann theorem does not apply for the qubit case, and choosing $\hat{\chi} \to \sigma_x$ and $\hat{\mu} \to \sigma_y$ will only satisfy (\ref{conditions}b,c) to first order in $\eta$. Hence the smallness of $\eta$ plays a crucial role in satisfying all three conditions.  Introducing $\etabar := \eta\, \textrm{mod} \, 2\pi $, we can expand $\mathcal{D}_{10}$ near $\etabar = 0$ (for general system dimension and measured observable $\hat{A}$ with eigenvalues $\{\alpha_i\}$) to first order:
\begin{subequations}
\begin{align}
\langle Q_j | Q_i \rangle & =\cos (\eta(\alpha_{j} - \alpha_i)) \approx 1, \\
\langle Q_{j}|\sigma_x| Q_i \rangle &= \sin(\eta(\alpha_{j} + \alpha_i)) \approx \etabar(\alpha_{j} + \alpha_i), \\
\langle Q_{j}|\sigma_y| Q_i \rangle &= - i \sin(\eta(\alpha_{j} - \alpha_i)) \approx -i \etabar(\alpha_{j} - \alpha_i).
\end{align}
\end{subequations}
Hence for fixed $\hat{A}$, choosing small enough $\etabar$ will allow the satisfaction of conditions (\ref{conditions}). However, the specific requirements for the three conditions are not all the same. Note that we require both cases of $\etabar^2(\alpha_{j} \pm \alpha_i)^2 \ll 1$. Thus, for a fixed $\etabar$, some weak values $\langle\hat{A} \rangle_f^W$ can be weakly measured, and others not; the breakdown of the AAV effect may manifest separately in the real and imaginary parts of the weak value (e.g., the sum may be large while the difference small).

\section{Weak Echoes} \label{sec:Weakechoes}

We have seen how a von Neumann interaction with small measurement parameter $\eta$ and post-selection can lead to the AAV effect. It turns out the AAV effect can be observed even outside the regime $\eta \to 0$. We saw in the previous section that, for qubit pointer, the three weak value conditions can be met when $\etabar = \eta\, \textrm{mod} \, 2\pi \to 0$. In this case, any measurement parameter which is near an integer multiple of $2\pi$ will lead to the AAV effect. This ought not be surprising since the measurement was a rotation which is itself periodic. Nevertheless, this emphasizes that one should think of the strength of a measurement not in terms of $\eta$, but rather in terms of pointer state distinguishability. %In other words, one should expect the AAV effect not just for small $\eta$, but for situations where $\mathcal{D}_{ij} \approx 1$.

It turns out that a modified AAV effect can manifest for the slightly weaker condition $|\mathcal{D}_{ij}|\to 1$. For the qubit pointer, substituting the approximate values for $\etabar = (\eta - \pi) \textrm{mod} \, 2\pi \ll 1$, making $\mathcal{D}_{10} \approx -1$, we find that the pointer shift is approximating the weak value
\begin{equation} \label{weakecho}
\frac{1}{2\etabar} \left\langle \sigma_x + i  \sigma_y \right\rangle_{B | A = f} \approx \frac{ \langle f| \sigma_z \hat{A} \rho  \sigma_z|f\rangle }{\langle f|\sigma_z \rho \sigma_z| f \rangle} = \langle \hat{A} \rangle^W_{\bar{f}},
\end{equation}
where $\bar{f}$ denotes a (potentially) different outcome to $f$, i.e., corresponding to quantum state $|\bar{f}\rangle = \sigma_z |f\rangle$. Note that the post-selection is still $f$, but the pointer shift is given by the weak value corresponding to $\bar{f}$. The explanation is simple: the $\pi$ phase-shift that leads to $\mathcal{D}_{10} = -1$ gets projected into the system, modifying the interference accordingly. This has a practical application. In a general-state direct measurement experiment, such as that of ref. \cite{Salvail:2013bo}, the post-selection had to be performed for all elements of the conjugate basis. This required a setup to projectively measure both diagonal and anti-diagonal polarization states. One could instead keep a single post-selection fixed in the setup, and use different values of $\eta$ to control the interference of the pointer states and thus cycle through the entire conjugate basis reading out the full Dirac distribution. 

\begin{figure}[htbp]
\begin{center}
\includegraphics[]{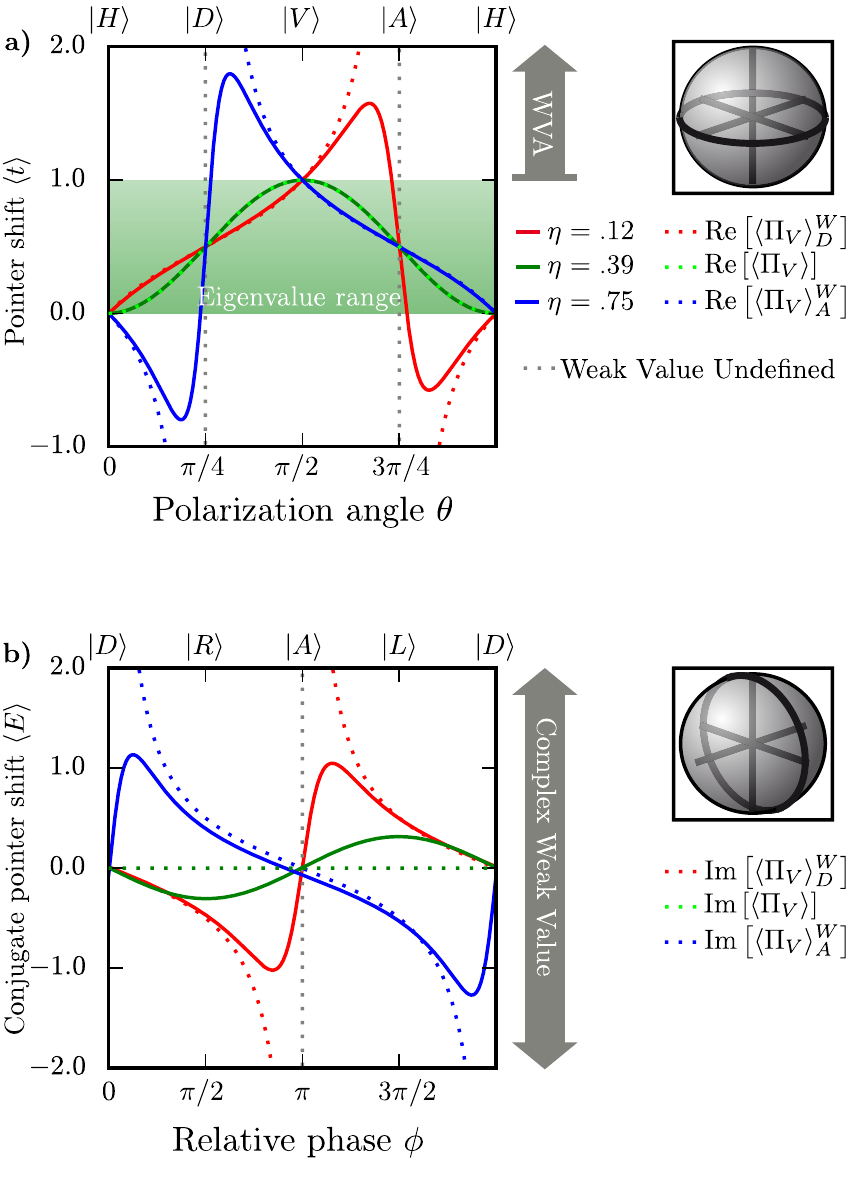}
\caption{\label{pointershifts_re_im} %I am the night.
Pointer shifts calculated according to equation (\ref{generalcase}), for the optical pulse hypothetical experiment; the measured system is the polarization degree of freedom of the light. The red, green, and blue solid curves represent the pointer shift for the three values of $\eta$ indicated by coloured discs in figure \ref{D_jis}. The dotted lines indicate the system quantities approximated by the correspondingly coloured pointer shift: weak value, expectation value, weak echo respectively. \textbf{a)} Canonical pointer shift $\langle t \rangle$ for input states around the equator of the Poincar\'{e} sphere, i.e., linearly polarized states parameterized by $|\psi\rangle = \cos \theta |0\rangle + \sin \theta |1\rangle$. The green region shows the eigenvalue bound of $[0,1]$, which the outcomes of strong measurements are restricted to, but which weak values can exceed; the weak and weak echo pointer shifts can exceed the bound, corresponding to weak value amplification (WVA). \textbf{b)} Conjugate pointer shift $\langle E \rangle$ for input states around a great circle of the Poincar\'{e} sphere, states parameterized by $|\psi\rangle = 1/\sqrt{2}( |0\rangle + \exp(i \varphi)  |1\rangle)$. In order to calculate these curves, the reduced strength $\etabar$ was used, taken to be the point where $\mathcal{D}_{10}$ crosses zero nearest the chosen strength, so that the pointer response $\langle t \rangle$ and $\langle E \rangle$ are approximately linear in $\etabar$ when expanded about this point, according to equations (\ref{conditions}b,c).
}
\end{center}
\end{figure}

If the pointer states are not periodic, we can still observe an AAV-like effect outside the regime $\eta \to 0$, using quasi-periodic or ``fringey'' pointers, like the optical pulse described above. Dressel and Jordan also investigated fringey pointers, by considering Hermite-Gaussian modes as a generalization of the usual Gaussian used in the weak value literature \cite{Dressel:2012ic}. As mentioned, for these continuous pointer states the second two conditions (\ref{conditions}b,c) are satisfied exactly, so the only remaining condition for the AAV effect is the pointer state distinguishability. The scheme then is to tune $\eta$ such that $\mathcal{D}_{10}$ nears a local maximum or minimum. Even if the curve will not reach unity in the neighbourhood, we can still observe AAV-like behaviour (amplification and conjugate pointer shift) in realistic scenarios. This is shown in figure \ref{pointershifts_re_im}, which shows exact calculations of the pointer shift using equation (\ref{generalcase}) for three values of $\eta$ for the optical pulse. The ratio of the pointer shift to the measurement parameter $\eta$ exceeds the eigenvalue range $[0,1]$ that limits the standard expectation value. In fact, it is possible for the weak value amplification for the weak echo to be greater than for the small-$\eta$ measurement (though reducing $\eta$ further in the latter case improves the approximation and hence amplification). Furthermore, the conjugate pointer shift, which reveals the imaginary part of the weak value, also exhibits weak value behaviour at this higher strength measurement. Interestingly, the strong measurement, for which $\mathcal{D}_{10} = 0$ (green), exhibits conjugate pointer shift. This can be attributed to the non-vanishing derivative of the post-selection probability with respect to measurement parameter $\eta$, supporting the claim that the conjugate pointer shift, in such a conditioned measurement, is related to the measurement back-action \cite{Dressel:2012dr}.

\section{Conclusions} \label{sec:Conclusions}
We have seen that weak values emerge in the usual sense, as the conditioned pointer shift, given three simple conditions on a subsystem of an entangled state and the operators used to characterize it. The most important condition is approximate pointer state indistinguishability. With this greater physical insight, valid for all measurement strengths, we can design new experiments based solely on the distinguishability of pointer states, and understand or interpret such experiments on their own terms in this framework.

Using this result, a new phenomenon is discovered and explained, whereby increasing the strength of the measurement can lead to the loss and recovery of weak-value behaviour. This means weak value amplification and conjugate pointer shift, the two useful aspects of weak values, are not limited to ``weak'' measurement, in the usual sense where the measurement strength approaches zero, but can be observed in a wider set of situations.

\begin{acknowledgements}
I am in great debt to J. Leach for helpful discussions during all stages of this work, and for hosting me at Heriot-Watt University where this work began. I am grateful to M. Malik for helpful discussions and giving the name of the weak echo. I thank G. Knee for his helpful comments on the manuscript.
\end{acknowledgements}

%\bibliographystyle{spphys}
%\bibliography{weak-echoes-refs}

\end{document}